\documentclass[a4paper,11pt]{article}
\pdfoutput=1 
\usepackage{jheppub} 
\usepackage{hyperref}
\usepackage{booktabs}
\usepackage[T1]{fontenc} 
\usepackage{siunitx} 
\usepackage{graphicx}%
\usepackage[dvipsnames]{xcolor}
\usepackage{dcolumn}
\usepackage{bm}
\usepackage[normalem]{ulem}
\usepackage{braket} 
\usepackage{amsmath}
\usepackage{cancel}
\usepackage{slashed}
\usepackage{multicol}
\usepackage[capitalise]{cleveref}
\usepackage{mathtools} 

\counterwithin*{equation}{section}
\allowdisplaybreaks

\newcommand{\nnl}{\nonumber \\}

\newcommand{\cO}{\mathcal{O}}

\newcommand{\cM}{{\mathcal M}}

\newcommand{\mpl}{M_{\rm Pl}}

\usepackage{etoolbox,siunitx}
\robustify\bfseries

\usepackage{subcaption}
\newcommand{\la}{\langle}
\newcommand{\ra}{\rangle}

\begin{document}

\title{
Gauge anomalies on shell and collinear factorization
}

\author{Edoardo Alviani and}
\author{Adam Falkowski }

\affiliation{Universit\'{e} Paris-Saclay, CNRS/IN2P3, IJCLab, 91405 Orsay, France}

\emailAdd{
alviani@ijclab.in2p3.fr,
adam.falkowski@ijclab.in2p3.fr
}

\abstract{
We revisit gauge anomalies from the purely on-shell perspective. We argue that violation of anomaly cancellation conditions manifests as a breakdown of collinear factorization. We explicitly construct one-loop 5-point amplitudes with graviton exchange displaying singular behavior in collinear limits that cannot be reconciled with factorization theorems.  In this approach, gravity serves as a universal probe of both abelian and non-abelian anomalies. 
}
\maketitle

\section{Introduction} 
\label{sec:INTR}

An anomaly in a quantum field theory (QFT) arises when the divergence of a classical symmetry current becomes non-zero due to quantum effects entering via one-loop diagrams~\cite{Adler:1969gk,Bell:1969ts}. 
While anomalous global symmetries may lead to interesting physical effects, local (gauge) symmetries must be free of anomalies for a QFT to be consistent~\cite{Bardeen:1969md}. 
The requirement of gauge anomaly cancellation imposes constraints on the allowed charges and representations of chiral fermions.

It is interesting to investigate how the gauge anomalies are reflected in the on-shell amplitude approach~\cite{Bern:1994cg,Bern:1996je,Britto:2005fq,Benincasa:2007xk}.  
In this framework, the basic building blocks are 3-particle amplitudes defined for complex on-shell external momenta, which are very constrained by the Poincar\'{e} invariance, locality, and little group covariance.   
Higher-point amplitudes are then obtained by bootstrapping the lower-point ones. 
This allows one to calculate scattering amplitudes in a straightforward and transparent fashion, which is especially advantageous when particles with spin $S \geq 1$ are involved.   
However, certain features of the on-shell approach appear to be at odds with the traditional notion of anomalies.  
One is that gauge symmetry is never introduced in the first place. 
Another is that on-shell 3-point amplitudes are tree-level exact, obscuring the link with the triangle anomalies entering via one-loop Feynman diagrams in the standard QFT calculations in four dimensions. 

Anomalies were discussed from the purely on-shell perspective in  Refs.~\cite{Huang:2013vha,Chen:2014eva}.  
 Using unitarity-based methods, the authors constructed the one-loop on-shell 4-gluon amplitude in four dimensions with chiral fermions in the loop. 
 They argued that the resulting amplitude may be at odds with locality when the standard anomaly cancellation conditions are not satisfied.    
 In detail, considering the  $t \to 0$ limit of the color-ordered four-gluon amplitude 
 $M[1^+ 2^- 3^+ 4^-]$, 
 the non-rational pieces required by unitarity produce $\frac{1}{t^2}$ and $\frac{1}{t}$ poles which are argued to be forbidden  by locality.
 In the case of a non-chiral  theory these can be eliminated by a suitable choice of the rational term, thus giving a method to fix part of the rational term from unitarity cuts. 
 However, for a chiral  theory the required rational term may introduce a new factorization channel, which is argued to be always inconsistent. 
 To avoid the offending terms one has to impose a condition $f^{abc} d^{ade}=0$ on the color factors,  which is precisely the box anomaly cancellation condition~\cite{Bilal:2008qx}. Ref.~\cite{AccettulliHuber:2021uoa} observed that the same reasoning can be applied in the case of abelian anomalies by considering 4-point amplitudes with photons and gravitons. 

In this paper we revisit the question of an on-shell formulation of anomalies. 
Our motivations are twofold.
First, we want to explore the interplay between gauge anomalies and gravity. 
The graviton -  the massless spin-2 messenger of gravitational interactions - must have a minimal coupling to all particles (including itself) whose magnitude is fixed by the  Planck scale.  
Therefore, it can serve as a  universal probe of the overall structure of the theory. 
Indeed, we will show that that amplitudes with external spin-1 vectors and a single graviton emission carry on-shell information about anomalies in both abelian and non-abelian theories.  
Second, we wish to put on a more solid footing the discussion about  incompatibility between  locality and unitarity for anomalous on-shell amplitudes. 
The arguments of Refs.~\cite{Huang:2013vha,Chen:2014eva} rely on the singular structure of on-shell 4-point amplitudes.  
However, for real momenta, the constraints due to 4-particle kinematics imply vanishing of the residues to the $u$-poles discussed above~\cite{unpublished}. 
For this reason we will turn to higher-point amplitudes, where  kinematics is less constraining.
In fact, the consequences of unitarity for the  leading behavior of loop amplitudes in the collinear limit  were worked out in Refs.~\cite{Bern:1995ix,Kosower:1999xi}, and these are formulated for $n$-point amplitudes with $n \geq 5$. 
For this reason, we will study certain on-shell 5-point amplitudes with graviton exchange, and we will show that their singular behavior is incompatible with the factorization theorem of Refs.~\cite{Bern:1995ix,Kosower:1999xi} when the theory contains gauge anomalies. 
In contrast to Refs.~\cite{Huang:2013vha,Chen:2014eva}, we use gravitational amplitudes to probe gauge anomalies, both abelian and non-abelian theories are investigated, and collinear factorization breakdown in 5-point amplitudes to pinpoint inconsistencies. 

This paper is organized as follows. 
In~\cref{sec:GRAV} we present the unitarity-based calculation of the one-loop 4-point scattering amplitude with 3 photons and 1 graviton, 
$\cM[1_\gamma^- 2_\gamma^- 3_\gamma^+ 4_h^+]$, in an abelian theory with chiral fermions. 
The result is proportional to the triangle anomaly cancellation condition and vanishes in an anomaly-free theory. 
We discuss the kinematic  structure of that amplitude in the case when anomalies do not cancel. 
We also summarize   analogous results for  amplitudes with photons replaced by gluons, as well as  for amplitudes with  3 gravitons and 1 vector boson. 
In \cref{sec:FACT} we review the requirements imposed by unitarity on the collinear limits of amplitudes, and we highlight the difficulty of applying these arguments to 4-point functions.
In~\cref{sec:BREAK} we move to discuss the 5-point amplitude with 3 photons and two chiral fermions. 
The singularity structure of that amplitude depends, via unitarity, on the previously calculated 
$\cM[1_\gamma^- 2_\gamma^- 3_\gamma^+ 4_h^+]$. 
We demonstrate this singularity structure cannot be reconciled with the factorization theorem of Refs.~\cite{Bern:1995ix,Kosower:1999xi}. 
We also  generalize these arguments to non-abelian theories and to mixed $U(1)$-gravitational anomalies.
\cref{sec:CONC} contains our conclusions and discusses possible future directions. 
\cref{app:INT} reviews the methods of the phase space integration needed to derive 1-loop amplitudes and the basis of 1-loop massless scalar integrals. 
Finally, in \cref{app:INT_B} we show an example of how collinear factorization works for a 5-point amplitude {\em not} proportional to an anomaly cancellation condition.

\textbf{Conventions}.
We work with the mostly minus Minkowski metric tensor $\eta_{\mu \nu} = (+,-,-,-)$. 
We use the spinor helicity formalism following the conventions of Ref.~\cite{Dreiner:2008tw}. 
The usual angle and square bracket notations is used to denote the holomorphic and anti-holomorphic spinors, such that 
$p_n \sigma  = \ket{n} [n|$. 
We suppress $\sigma$ matrices when sandwiched between spinors, e.g. $\bra{1}p_2 |3] \equiv \bra{1}p_2 \sigma |3]$.  
On-shell amplitudes are written in the all-incoming convention.   
We define the Mandelstam invariants $s_{nm} = (p_n + p_m)^2$ and for 4-point amplitudes we use  
$s,t,u=(p_1+p_{2,3,4})^2$.

\section{Gravity as a probe for anomalies} 
\label{sec:GRAV}

In this section we compute the one-loop four-point amplitudes which will be needed to probe anomalies later on. 
We compute these amplitudes up to rational terms using locality and unitarity. Namely, starting from three-point amplitudes, we use unitarity to reconstruct tree-level four-point amplitudes, and then calculate unitarity cuts to reconstruct the loop-amplitudes of interest.

Consider first a theory with a photon ($\gamma$) a graviton ($h$), and $N$ chiral fermions ($\psi_n$). 
The building blocks of our theory are the 3-point amplitudes: 
\begin{align}
\label{eq: 3-points LH abelian}
    &\cM[1_\gamma^- 2_{\psi_n}^- 3_{\bar\psi_n}^+]=
    \sqrt{2}e Q_n \frac{\la12\ra^2}{\la23\ra}  
    , \qquad \qquad \cM[1_\gamma^+ 2_{\psi_n}^- 3_{\bar\psi_n}^+]=
    -\sqrt{2}e Q_n \frac{[13]^2}{[23]}
    , \notag \\
&\cM[1_h^- 2_{\psi_n}^- 3_{\bar\psi_n}^+]=
-\frac{1}{\mpl}
\frac{\la12\ra^3 \la13\ra}{\la23\ra^2}
    , \; \; \; \; \; \; \; \cM[1_h^+ 2_{\psi_n}^- 3_{\bar\psi_n}^+]= 
    \frac{1}{\mpl} \frac{[13]^3 [12]}{[23]^2}. 
    \\
    &\cM[1_\gamma^- 2_{\psi_n}^+ 3_{\bar\psi_n}^-]=
    - \sqrt{2}e Q_n \frac{\la13\ra^2}{\la23\ra}  
    , \qquad \quad \; \cM[1_\gamma^+ 2_{\psi_n}^+ 3_{\bar\psi_n}^-]=
    \sqrt{2}e Q_n \frac{[12]^2}{[23]}
    , \notag \\
    &\cM[1_h^- 2_{\psi_n}^+ 3_{\bar\psi_n}^-]=
    \frac{1}{\mpl}\frac{\la13\ra^3 \la12\ra}{\la23\ra^2}
    , \quad \, \; \; \; \; \; \; \cM[1_h^+ 2_{\psi_n}^+ 3_{\bar\psi_n}^-]= - \frac{1}{\mpl} \frac{[12]^3 [13]}{[23]^2},
\end{align}
where the first (second) equations describes a left- (right-) handed fermion\footnote{We define a left-hand fermion as the case when the particle $\psi_n$ has negative helicity whereas the anti-particle $\bar \psi_n$ has positive helicity, and the right-handed fermion as the opposite case.
}.
Using the unitarity factorization, we then find the needed tree-level four-point amplitudes reconstructing them from their residues as 
\begin{align}
&\cM^{(0)}[1_\gamma^- 2_\gamma^+ 3_{\psi_n}^- 4_{\bar\psi_n}^+]=
\la13\ra [24] \la 1|p_3|2] \bigg[\frac{2 e^2 Q_n^2}{t u} - \frac{1}{\mpl^2 s}\bigg]
, \notag \\
&\cM^{(0)}[1_\gamma^- 2_h^+ 3_{\psi_n}^- 4_{\bar\psi_n}^+]=
\frac{\sqrt{2} e Q_n}{\mpl s t u} \la13\ra [24] [12] \la1|p_3|2]^2
, \\
&\cM^{(0)}[1_\gamma^- 2_\gamma^+ 3_{\psi_n}^+ 4_{\bar\psi_n}^-]=
\la14\ra [23] \la1|p_3|2] \bigg[\frac{2 e^2 Q_n^2}{t u} - \frac{1}{\mpl^2 s}\bigg]
, \notag \\
&\cM^{(0)}[1_\gamma^- 2_h^+ 3_{\psi_n}^+ 4_{\bar\psi_n}^-]=
\frac{\sqrt{2} e Q_n}{\mpl s t u} \la14\ra [23] [12] \la1|p_3|2]^2
.\end{align}
The tree-level amplitudes are determined up to contact terms, which correspond to higher-dimensional operators and play no role in the discussion of anomalies. 

We now have all the ingredients to compute the one-loop amplitude 
$\cM^{(1)}[1_\gamma^- 2_\gamma^- 3_\gamma^+ 4_h^+]$ using unitarity bootstrap.
We will do it using only two-particle cuts, which give us the discontinuity of the amplitude across its branch cuts: 
$\text{Disc}_x f(x) = 
f(x+i\epsilon)
-f(x-i\epsilon)$. 
The branch cuts of our amplitude are along the $t$ and $u$ channels, and correspond to pair production of chiral fermions. 
By unitarity, the $t$-channel discontinuity is related to the tree-level amplitudes by  
\begin{align}
\text{Disc}_t \cM^{(1)}[1_\gamma^- 2_\gamma^- 3_\gamma^+ 4_h^+] =  & 
-i \sum_n \int \text{d}\Pi_{XY} \cM[1_\gamma^- 3_\gamma^+ (-Y)_{\psi_n}^{s_n} (-X)_{\bar\psi_n}^{-s_n}] 
\cM[2_\gamma^- 4_h^+ X_{\psi_n}^{s_n} Y_{\bar\psi_n}^{-s_n}]
, \end{align}
where $s_n=-(+)$ for left- (right-) handed fermions,
and $d\Pi_{XY}$ is the two-body phase space element for the fermion pair in the loop.
Plugging in the tree-level amplitudes and using the techniques summarized in \cref{app:INT} to integrate over the phase space, we can express the right-hand side a as a sum of $t$-channel discontinuities of the box, triangle and bubble scalar integrals defined in \cref{eq:INT_scalar-integrals}:
\begin{align}
\text{Disc}_t \cM^{(1)}[1_\gamma^- 2_\gamma^- 3_\gamma^+ 4_h^+] =  & 
 - \bigg [ \sum_{n=1}^N  s_n  Q_n^3  \bigg ] 
     {  \sqrt 2 e^3  \over  \mpl} 
{ \langle 12 \rangle^2 [34]^2
[4| p_1   p_2 | 4 ]
\over  s^4}  
 \nnl \times & 
 \text{Disc}_t \bigg\{  
4  s  I_\circ^t  
 + t (t - u)  \big [   
 u I_{\Box}^{tu} 
 - 2 I_{\triangle}^t  \big ]  \bigg \}
. \end{align}
The $u$-channel discontinuity 
$\text{Disc}_u \cM^{(1)}$ is obtained from the above by  
$1 \leftrightarrow 2$. 
Reconstructing the amplitude and replacing the scalar integrals with the analytic expressions in \cref{eq:INT_BubbleTriangleBoxAnalytic} we finally obtain
\begin{align}
\label{eq:FACT_M3gamma1h}
\cM^{(1)} \big[ 1_\gamma^- 2_\gamma^- 3_\gamma^+ 4_h^+ \big] = & 
M_{3\gamma h}  
+ R_{3\gamma h},
 \notag \\  
  M_{3\gamma h}  \equiv & 
 -\bigg[\sum_{n=1}^N s_n Q_n^3 \bigg] 
\frac{\sqrt{2} e^3}{16 \pi^2 \mpl} 
\langle 12 \rangle^2 [34]^2
[4| p_1   p_2 | 4 ] F_{3\gamma h}(s,t) ,
 \notag \\ 
F_{3\gamma h}(s,t)  \equiv  &
{1 \over s^3} 
\bigg \{  4 \log\bigg(\frac{t}{u}\bigg) 
+ {t - u \over s} 
\bigg [  \log\bigg(\frac{t}{u}\bigg)^2 + \pi^2 \bigg ] 
\bigg \}|_{u=-s-t}
,\end{align}
where $\log(t/u) \equiv \log(-t) - \log(-u)$ and   $R_{3\gamma h}$ is a rational term that cannot be reconstructed via unitarity cuts in four dimensions. 
Notice that the amplitude has no UV or IR divergencies.  
Furthermore, as already noticed in Ref.~\cite{AccettulliHuber:2021uoa}, it is proportional to the $U(1)^3$ anomaly. 
In particular, it is zero in a non-chiral theory, where for each left-handed fermion there is a right-handed one canceling its contribution. 
That is of course ensured by Furry's theorem \cite{Furry:1937zz} as a consequence of $C$ conservation.\footnote{%
For the same reason, scalar loop contributions to that amplitude vanish. 
}

In the remainder of this section we quote without derivation the results for other relevant one-loop 4-point amplitudes with chiral fermions in the loop, which can be obtained by the same techniques.  
First, we have the amplitude with a single photon and 3 gravitons: 
 \begin{align}   
\label{eq:FACT_Mhhha-mmpp} 
\cM^{(1)}[1_h^- 2_h^- 3_h^+ 4_\gamma^+]  = & 
M_{3h\gamma} + R_{3h\gamma},
\nnl 
M_{3h\gamma} \equiv & 
\bigg [ \sum_{n=1}^N  s_n  Q_n  \bigg ] 
   { \sqrt 2 e \over   32 \pi^2 \mpl^3}  
   { \langle 12 \rangle^4
   [34]^2  [3| p_1 p_2|3]
     \over s^3} 
{t u \over s^2}  
F_{3h\gamma}(s,t),
\nnl 
F_{3h\gamma}(s,t) \equiv & 
\bigg \{ {t^2 - 10 t u + u^2 \over 3 t u }  \log \bigg ( { t \over u } \bigg ) 
-  {t-u \over s} \bigg [  \log^2 \bigg ( { t \over u } \bigg )   + \pi^2 \bigg ]  
\bigg \}|_{u=-s-t}
.   \end{align}   
Much as its 3-photon-1-graviton counterpart, this vanishes in a non-chiral theory by Furry's theorem. 
In a chiral theory it is proportional to the mixed $U(1)$-gravitational anomaly~\cite{AccettulliHuber:2021uoa}. 

Finally, consider a non-abelian gauge theory coupled to gravity and to 
$n_R^{\cal R}$($n_L^{\cal R}$) right(left)-handed massless fermions in the representation ${\cal R}$.
The one-loop 3-gluon-1-graviton amplitude then reads 
\begin{align}
\label{eq:FACT_M3g1h}
&\cM^{(1)}\big[1_a^- 2_b^- 3_c^+ 4_h^+\big] =M_{3gh}+R_{3gh} 
, \notag \\
&M_{3gh}=
\frac{ g^3}
{16 \sqrt{2} \pi^2 \mpl} 
\langle 12\rangle^2 [34]^2 [4|p_1 p_2|4] 
\bigg [
i n_+^{\cal R}  C_{\cal R}  f^{abc} G_{3gh}(s,t)  
+ n_-^{\cal R} d_{\cal R}^{abc}F_{3\gamma h}(s,t) \bigg ] 
, \notag \\
& G_{3g h}(s,t)  \equiv \frac{1}{s^3} \bigg\{
\frac{ t^2 + u^2 }{s^2}
\bigg[\log^2\bigg(\frac{t}{u}\bigg) + \pi^2 \bigg] 
+ \frac{4 s^2}{3 t u }  \bigg(\frac{1}{\varepsilon} + 2\log\mu + 2 \bigg) 
\notag\\
&\qquad \qquad \quad+ 2 \log(-t) \frac{2s^2 + 3st + 6t^2}{3st} + 2 \log(-u) \frac{2s^2 + 3 s u + 6u^2 }{3su}  \bigg\}|_{u=-s-t}
 .\end{align}
where 
$ f^{abc} = 
- {i \over C_{\cal R} }   \text{Tr}\big[[T^a,T^b]T^c\big]$ with $C_{\cal R} = 1/2$ for the fundamental representation, and 
$d_{\cal R}^{abc} = \text{Tr}\big[\{T^a,T^b\}T^c\big]$.  Furthermore, $\varepsilon$ is a dimensional regulator defined as 
$d=4-2\varepsilon$, $\mu$ is the renormalization scale, and 
$n_\pm^{\cal R} = n_R^{\cal R} \pm n_L^{\cal R}$.
Notice that the kinematic function multiplying $d_{\cal R}^{abc}$ coincides with the one in \cref{eq:FACT_M3gamma1h}, and with the one occurring in the 4-gluon amplitude in Ref.~\cite{Huang:2013vha}.  

\section{Collinear factorization and unitarity}
\label{sec:FACT}

\begin{figure}
    \centering
\includegraphics[width=0.9\linewidth]{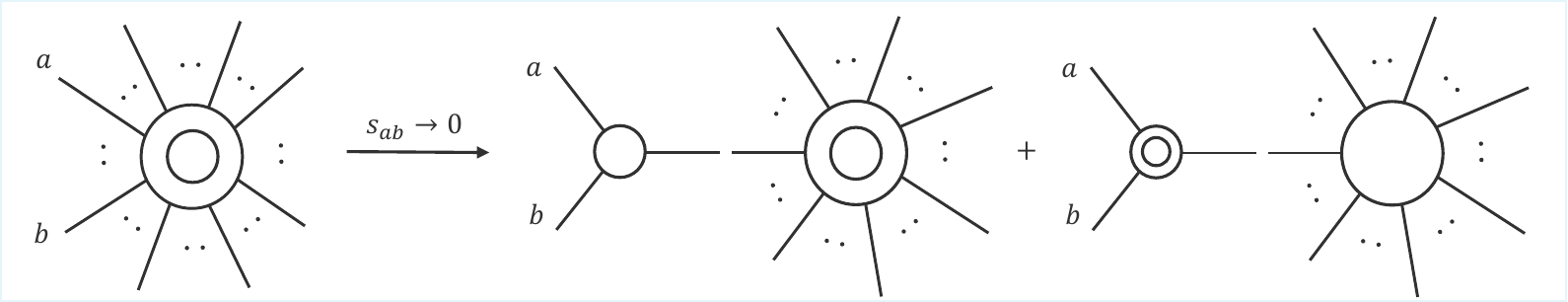}
    \caption{Illustration of the collinear factorization theorem in \cref{eq:FACT_Factorization-loop}. }
    \label{fig:FACT_diagrams}
\end{figure}

Amplitudes in gauge and gravitational theories display a universal behavior in the limit where two momenta become collinear, $p_a = z P$, $p_b = (1-z)P$, 
$P^2=0$. 
At tree level, amplitudes factorize into the so-called {\em splitting function} and lower-point amplitudes: 
\begin{align}
\label{eq:FACT_Factorization-tree}
&\cM_n^{(0)} \big[ \dots a^{\lambda_a} b^{\lambda_b} \dots  \big]  \xrightarrow{a\|b}
\sum_{x,\lambda=\pm}
 \text{Split}^{(0)}_{x^{-\lambda}}\big[ a^{\lambda_a} b^{\lambda_b} \big] 
\cM_{n-1}^{(0)}\big[\dots P^\lambda \dots\big] 
 ,\end{align}
 where one should sum over the species $x$ and the helicities $\lambda$ of the splitting state with momentum $P$.
The splitting functions  $\text{Split}^{(0)} = - {\rm lim}_{P^2\to 0}
{\cM[a b (-P)] \over P^2}$~\cite{Badger:2023eqz} encode the singular behavior of amplitudes in the collinear limit. 
They are  purely rational functions of spinor brackets, and they are well known for theories of interest~\cite{Mangano:1990by,Bern:1998zm,Bern:1998sv}. 
In gauge theories, they display a square-root singularity, $\text{Split} \sim {1 \over \sqrt {s_{ab}} }$, while  
for gravitational theories they behave as $\text{Split} \sim s_{ab}^0$.\footnote{%
The ``singular" behavior in the gravitational case consists in $\text{Split} \sim {\langle a b \rangle \over  [ab]}$ or inverse.} 
Other than this asymptotics, the precise form of the splitting functions will not be important for our discussion. 

At one loop, the factorization theorem is generalized as~\cite{Bern:1995ix,Kosower:1999xi}
\begin{align}
\label{eq:FACT_Factorization-loop}
\cM_n^{(1)} \big[ \dots a^{\lambda_a} b^{\lambda_b} \dots  \big]  
\xrightarrow{a\|b} & 
\sum_{x,\lambda=\pm} \bigg\{ \text{Split}^{(0)}_{x^{-\lambda}}\big[ a^{\lambda_a} b^{\lambda_b} \big] 
\cM_{n-1}^{(1)}\big[\dots 
P^\lambda \dots\big] 
\nnl & 
+ \text{Split}^{(1)}_{x^{-\lambda}}\big[ a^{\lambda_a} b^{\lambda_b} \big] \cM_{n-1}^{(0)}\big[\dots P^\lambda  \dots\big] \bigg\}
 ,\end{align}
where the $(0)$ and $(1)$ superscripts describe tree-level and one-loop pieces, respectively. 
Compared to \cref{eq:FACT_Factorization-tree}, this also involves the one-loop splitting functions
$\text{Split}^{(1)}$. 
These may be more complicated and include dependence on the dimensional regulator or non-rational functions of kinematical parameters~\cite{Bern:1994fz,Kosower:1999rx}. 
Nevertheless they retain universality, in the sense that they only depend on the collinear kinematics, and they have the same singularity structure as the tree-level ones, up to eventual logarithms. 
Moreover, for gravitational theories  $\text{Split}^{(1)}$ vanishes, that is to say,  the gravitational splitting functions are tree-level exact~\cite{Bern:1998sv}. 
Thanks to these universal features, collinear factorization is a useful tool to verify the compatibility of concrete amplitudes with fundamental principles of QFT such as locality and unitarity.

With this mind, we will study the collinear properties of on-shell amplitudes in theories where the anomaly cancellation condition is not satisfied. 
We will argue, however, that the ``anomalous" 4-point amplitudes calculated in \cref{sec:GRAV} by themselves do not violate collinear factorization. 
Consider the 3-photon-1-graviton amplitude
in \cref{eq:FACT_M3gamma1h}.  
For small $s$,  the $F_{3\gamma h}$ function behaves as
 \begin{align} 
F_{3\gamma h}(s,t)   = & 
{2 \over s^2 t } + \cO(s^{-1}) 
, \end{align}
while it has only logarithmic singularities for $t$ or $u$ approaching zero. 
It follows that, in the $1\|2$ collinear limit, the non-rational part of the 3-photon-1-graviton one loop amplitude behaves as 
\begin{align} 
 M_{3\gamma h} \sim &  
 \sqrt{|s|}
, \end{align} 
which has a smooth limit as $s\to 0$. 
This is in line with \cref{eq:FACT_Factorization-loop} assuming that, as is the case in QED, 3-photon splitting functions vanish.


The situation is similar in a non-abelian gauge theory for the 3-gluon-1-graviton amplitude in \cref{eq:FACT_M3g1h}. 
The potentially anomalous part (proportional to $d^{abc}$) coincides, apart for some irrelevant prefactors, with that of \cref{eq:FACT_M3gamma1h}. 
By the same token that part has a smooth limit as  
$\sqrt{|s|}$ in the $1\|2$ collinear limit~\cite{unpublished}.
The non-anomalous piece (proportional to $f^{abc}$) scales differently. 
Using
 \begin{align} 
G_{3g h}(s,t)   = & 
-{2 \over s^3 } 
+ \cO(s^{-2})
,\end{align}
one sees that it scales as $1/\sqrt{|s|}$. 
Both of these scalings are in line with \cref{eq:FACT_Factorization-loop} and the presence of a one-loop gluon splitting function proportional to $f^{abc}$.

\begin{figure}
    \centering
\includegraphics[width=0.9\linewidth]{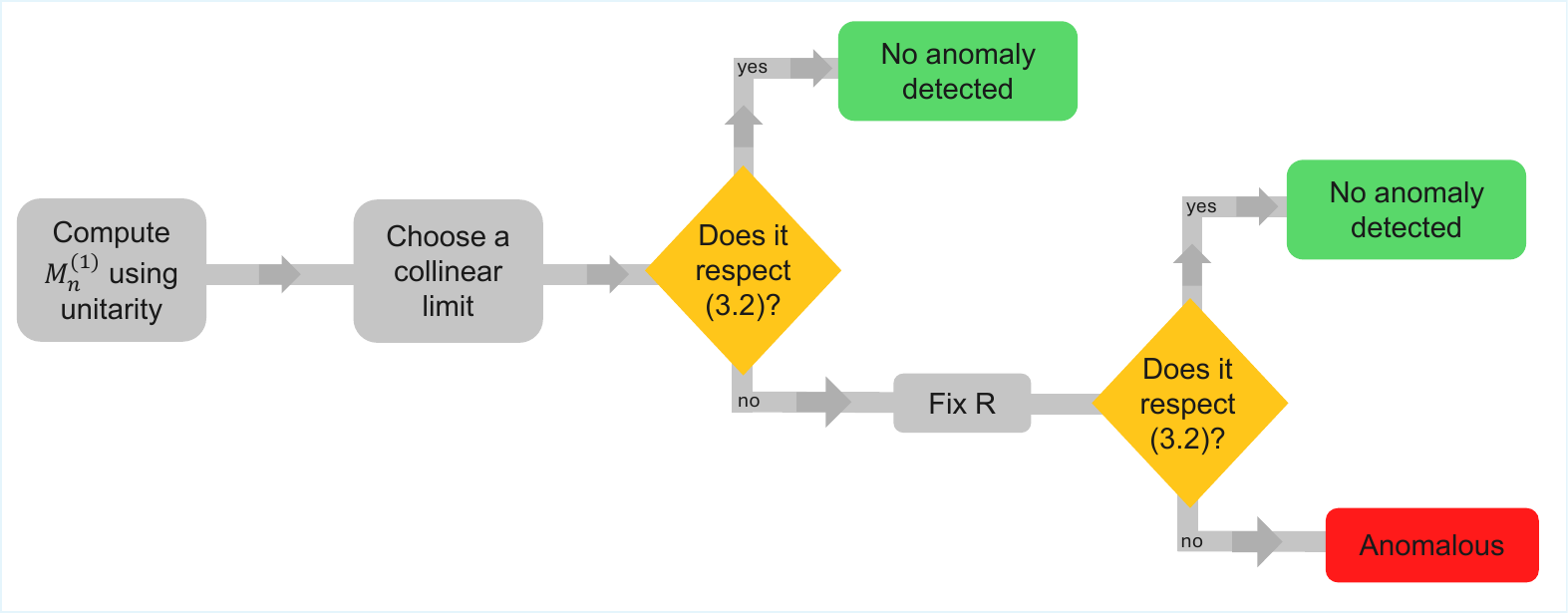}
\caption{%
Sketch of our algorithm to detect anomalies in on-shell 5-point amplitudes using the collinear factorization theorem of \cref{eq:FACT_Factorization-loop}.   }
    \label{fig:FACT_algorithm}
\end{figure}

One concludes that 4-point amplitudes, even those proportional to anomalies, do not display pathological collinear behavior. 
In the following, we will turn to higher-point amplitudes, where kinematics is less constraining and allows for more flexibility in the study of singular limits. 
In fact, proofs of collinear factorization at one loop require the number of external particles to be $n\geq5$~\cite{Bern:1995ix,Kosower:1999rx}.
To be specific, we will study certain one-loop 5-point amplitudes obtained by bootstrapping the one-loop 4-point amplitudes calculated in \cref{sec:GRAV}. 
With the 5-point amplitudes at hand, we will study their collinear limits to detect pathologies due to anomalies.
If factorization fails, we will try to restore \cref{eq:FACT_Factorization-loop} by a judicious choice of the rational term. 
When that is not possible, the theory is deemed anomalous. 
Our algorithm is sketched in \cref{fig:FACT_algorithm}. 

\section{Gauge anomalies as collinear factorization breakdown} 
\label{sec:BREAK}

There are several 5-point amplitudes where factorization breakdown in an anomalous theory can be demonstrated. Consider, for example, the one-loop amplitude 
$\cM^{(1)} [1_\gamma^- 2_\gamma^- 3_\gamma^+ 4_{\psi_k}^- 5_{\bar \psi_k}^+]$, where $\psi_k$ is one of the chiral fermions in the theory (here chosen to be left-handed for concreteness).\footnote{%
The fermion $\psi_k$ may be one contributing to the 3-photon-1-graviton amplitude in \cref{eq:FACT_M3gamma1h}, but it could also be ``dark", that is to say $Q_k = 0$, which would simplify the structure of the considered 5-point amplitude.}  
By unitarity, it contains a contribution from the graviton exchange
with the residue proportional to the anomalous amplitude in \cref{eq:FACT_M3gamma1h}: 
\begin{align} & 
\cM^{(1)} [1_\gamma^- 2_\gamma^- 3_\gamma^+ 4_{\psi_k}^- 5_{\bar \psi_k}^+]  \supset 
- { \cM ^{(1)} [ 1_\gamma^- 2_\gamma^- 3_\gamma^+ K_h^+] 
\cM[(-K)_h^- 4_{\psi_k}^- 5_{\bar \psi_k}^+ ]
\over s_{45} }
,  \end{align} 
where $K = p_4 + p_5$ and $s_{45} = K^2$. 
This equation can be viewed a consequence of collinear factorization, given the definition of a tree-level splitting function below \cref{eq:FACT_Factorization-tree} and that $\cM^{(0)}\big[ 1_\gamma^- 2_\gamma^- 3_\gamma^+ 4_h^+]=0$. 
Plugging in \cref{eq:FACT_M3gamma1h}, as well as the minimal coupling of gravity to spin-1/2 fermions, one obtains   
\begin{align}
\label{eq:BREAK_M3gamma2psi} 
\cM^{(1)} [1_\gamma^- 2_\gamma^- 3_\gamma^+ 4_{\psi_k}^- 5_{\bar \psi_k}^+]  \supset  & 
M_{3\gamma2\psi} 
+ R'_{3\gamma 2 \psi},
\nnl 
M_{3\gamma2\psi} =    & 
  \bigg[\sum_{n=1}^N s_n Q_n^3 \bigg] 
\frac{\sqrt{2} e^3}{16 \pi^2 \mpl^2   }
  { \langle 45 \rangle   \over [45] }
 \langle 12 \rangle^3  [15][24] [35]^2 
 F_{3\gamma h}(s_{12}, s_{13})  , 
\nnl 
R'_{3\gamma 2 \psi} =  &
{ R_{3\gamma h}[ 1_\gamma^- 2_\gamma^- 3_\gamma^+ K_h^+] \over s_{45}} 
{ \langle K 4 \rangle^3
\langle K 5 \rangle \over \mpl \langle 4 5 \rangle^2}
.   \end{align} 
One subtlety here is that,  for $s_{45} \to 0$, $M_{3\gamma2\psi} \sim  {\langle 45 \rangle   \over [45] }  \sim s_{45}^0$.
Therefore it is not enhanced and other terms in the amplitude can be numerically of  similar order for small $s_{45}$. 
In fact, this very behavior is enforced by factorization theorems, due to the corresponding gravitational splitting function behaving as  $\text{Split} \sim {\langle 45 \rangle   \over [45] }$. 
Nevertheless, $M_{3\gamma2\psi}+R'_{3\gamma2\psi}$ is distinct from other terms in the 5-point amplitude and can be studied in isolation, in the sense that there cannot be cancellation of potentially pathological terms between this and other parts of the amplitude.
One argument is that when rotating, say, $p_5$ around $p_4$ the factor 
$\langle 45 \rangle/[45]$ picks up a large phase proportional to the angle of rotation~\cite{Bern:1998sv} unlike other terms not included in $M_{3\gamma2\psi}+R'_{3\gamma 2 \psi}$.
Furthermore, the charge dependence of $M_{3\gamma2\psi}$ is not shared by any other term in the one-loop amplitude that could develop a singularity as $s_{45} \to 0$.
In particular, such other terms would be zero if we consider scattering of a dark fermion with $Q_k =0$. 

Now take the collinear limit $p_1 = z P$, $p_2 = (1- z) P$ of 
$M_{3\gamma2\psi}$  in~\cref{eq:BREAK_M3gamma2psi}:
\begin{align}
\label{eq:BREAK_M3gamma2psi-1|2} 
 & 
M_{3\gamma2\psi} \xrightarrow{1\| 2} 
     \bigg[\sum_{n=1}^N s_n Q_n^3 \bigg] 
\frac{\sqrt{2} e^3}{8 \pi^2 \mpl^2   }
  { \langle 45 \rangle   \over [45] }
 [P4] [P5] [35]^2 
{\sqrt{z(1-z)}  \over s_{13} } {\langle 1 2 \rangle^3 \over s_{12}^2 } 
\sim {1 \over  \sqrt{s_{12} }}
.  \end{align} 
This has a square root singularity, which is seemingly similar to collinear behavior in usual gauge theories. 
However, in this case it is  impossible to describe this  singularity by a factorization theorem with a universal splitting function.  
Indeed, we would need to factorize \cref{eq:BREAK_M3gamma2psi-1|2} using a splitting function for a graviton or a photon splitting into 2 photons. 
Starting with the former option, 
$\text{Split}_{h^\lambda}^{(0)}\big[ a_\gamma^{\lambda_a} b_\gamma^{\lambda_b} \big]  \sim s_{ab}^0$, therefore it cannot produce a square root singularity, whereas $\text{Split}_{h^\lambda}^{(1)}\big[ a_\gamma^{\lambda_a} b_\gamma^{\lambda_b} \big] = 0$ because gravitational splitting functions are tree-level exact. 
Moving to the latter option,  
$\text{Split}_{\gamma^\lambda}^{(0)}\big[ a_\gamma^{\lambda_a} b_\gamma^{\lambda_b} \big] = 0$  because photon has no self-interactions at tree level. 
Finally, even a pathological solution to allow for a non-zero 3-photon one-loop splitting function does not offer a way out in this factorization channel. 
One can show that it is impossible to find a universal 
$\text{Split}^{(1)}_{\gamma^\lambda}\big[ a_\gamma^{\lambda_a} b_\gamma^{\lambda_b} \big]$
that would fit with \cref{eq:BREAK_M3gamma2psi-1|2} and depend only on the collinear kinematics.\footnote{%
An ansatz 
$\text{Split}^{(1)}_{\gamma^+}\big[ 1_\gamma^- 2_\gamma^- \big]
 \cM^{(0)} [P_\gamma^- 3_\gamma^+ 4_\psi^- 5_{\bar \psi}^+] $ leads to the splitting function depending on the momenta $p_4$ and $p_5$. 
}

To save factorization, the ${1 \over \sqrt{s_{12}}}$ singularity in \cref{eq:BREAK_M3gamma2psi-1|2} has to be canceled by a  contribution from $R'_{3\gamma h}$, which originates from the rational term $R_{3\gamma h}$ in \cref{eq:FACT_M3gamma1h}. 
A possible choice is  
\begin{align} 
R_{3\gamma h} \big[ 1_\gamma^- 2_\gamma^- 3_\gamma^+ 4_h^+ \big]  = & 
  -      \bigg[\sum_{n=1}^N s_n Q_n^3 \bigg] 
\frac{\sqrt{2} e^3}{16 \pi^2 \mpl   }
{ \langle 12 \rangle^2 [34]^2 [4| p_1   p_2 | 4 ]  (t-u)    \over s^2  t u   } 
, \nnl 
R'_{3\gamma 2 \psi} \big[ 1_\gamma^- 2_\gamma^- 3_\gamma^+ 4_{\psi_k}^- 5_{\bar \psi_k}^+ \big] = & 
   \bigg[\sum_{n=1}^N s_n Q_n^3 \bigg] 
    \frac{\sqrt{2} e^3}{32 \pi^2 \mpl^2   } 
      {\langle 4 5 \rangle \over [45] }  
      \langle 12 \rangle^3  [35]^2   \big ( [15] [24 ]  + [25] [14] \big ) 
{ s_{13} - s_{23}   \over s_{12}^2 s_{13} s_{23}  }  
.  \end{align}  

At this point $M_{3\gamma 2 \psi} + R'_{3\gamma 2 \psi}$ is regular in the $1 \| 2$  collinear limit. 
But now in the $1 \| 3$  limit, $p_1 = z P$, $p_3 = (1-z)P$, the cut part of the 5-point amplitude is regular (up to logarithms) while $R'_{3\gamma 2 \psi}$  behaves as 
\begin{align} 
\label{eq:BREAK_Rprime-1|3}
R'_{3\gamma 2 \psi} \xrightarrow{1\|3} &  
  \bigg[\sum_{n=1}^N s_n Q_n^3 \bigg] 
    \frac{\sqrt{2} e^3}{16 \pi^2 \mpl^2   } 
    (1-z) 
 {     \langle  24  \rangle^2  \over  \langle 4 5 \rangle  }
 {   s_{24}     \over   s_{13}   } 
 \sim {1 \over s_{13}} 
.  \end{align}  
Trying to interpret this singularity in the light of the factorization theorem of \cref{eq:FACT_Factorization-loop}, one ends up with a non-universal splitting faction with a pathological singular behavior ($1/s_{xy}$  instead of $1/\sqrt{s_{xy}}$). 
It is not possible to amend that by adjusting rational terms while maintaining the 
$1\leftrightarrow 2$ Bose symmetry of the amplitude. Moreover, new particles in the spectrum that might contribute to this factorization channel cannot solve the problem either.
We conclude that  saving $1 \| 2 $ collinear factorization leads to another pathology in the $1 \| 3 $  (and, analogously $2 \| 3 $)  collinear limit. 
The only way to avoid a breakdown of factorization is to postulate the $U(1)^3$ anomaly cancellation condition 
 \begin{align} 
\sum_{n=1}^N s_n Q_n^3  = 0 
,   \end{align} 
such that the one-loop 3-photon-1-graviton amplitude vanishes. 

\vspace{1cm}

The above discussion of abelian $U(1)^3$ gauge anomalies generalizes with little changes to the non-abelian case. 
Starting with the one-loop 3-gluon-1-graviton amplitude in~\cref{eq:FACT_M3g1h}, we consider, for example, the 5-point amplitude $\cM[ 1_a^- 2_b^- 3_c^+ 4_h^- 5_h^+]$.\footnote{%
The argument would be very similar if we considered $\cM[ 1_a^- 2_b^- 3_c^+ 4_{\psi_k}^- 5_{\bar \psi_k}^+]$, similarly as in the abelian case; 
likewise, in the abelian case we could have demonstrated factorization breakdown using $\cM[ 1_\gamma^- 2_\gamma^- 3_\gamma^+ 4_h^- 5_h^+]$ as an example.
}
For $s_{45}\to 0$, this amplitude contains a contribution from the graviton exchange  
\begin{align} 
\label{eq:BREAK_M3g2h}  
\cM^{(1)} [1_a^- 2_b^- 3_c^+ 4_h^- 5_h^+]  \supset  & 
- { \cM ^{(1)} \big[ 1_a^- 2_b^- 3_c^+ K_h^+ \big ] 
\cM^{(0)} [(-K)_h^-  4_h^- 5_h^+ \big]  \over s_{45} } 
=
M_{3g2h} + R'_{3g2h} 
,\nnl   
 M_{3g2h} = &
-  \frac{g^3}{16 \sqrt{2} \pi^2 \mpl^2} { \langle  4  5\rangle \over [45]}
 { \langle 12\rangle^2    [35]^4 [5| p_1 p_2 |5] \over [34]^2 } 
 \nnl \times  & 
 \bigg [ 
i n_+^{\cal R} C_{\cal R} f^{abc}  G_{3gh}(s_{12},s_{13})  
+ n_-^{\cal R}  d_{\cal R}^{abc}  F_{3\gamma h} (s_{12},s_{13})  \bigg ], 
\nnl 
R'_{3g2h} =  &
  { R_{3 g h} [ 1_a^- 2_b^- 3_c^+ K_h^+] \over s_{45} } 
  {  \langle K 4 \rangle^6 \over  \mpl \langle K 5 \rangle^2 \langle 4 5 \rangle^2  } 
,    \end{align}
where $K = p_4 + p_5$. 
In the limit where the 4-point rational term $R_{3 g h}$ vanishes, the 5-point amplitude in  \cref{eq:BREAK_M3g2h} runs into problems with collinear factorization in the $1\|2$ channel. 
In order to get rid of the offending singularity one needs to adjust $R_{3 g h}$ as  
\begin{align}
R_{3gh}=
\frac{g^3}{16 \sqrt{2} \pi^2 \mpl} 
\langle 12\rangle^2 [34]^2 [4|p_1 p_2|4] 
\bigg [ 
i n_+^{\cal R} C_{\cal R} f^{abc}  
{2 \over s^3 } 
+ n_-^{\cal R} d_{\cal R}^{abc}  {t - u \over s^2 t u }  \bigg ] 
, \end{align} 
such that the 5-point amplitude contains 
\begin{align} 
\label{eq:BREAK_M3g2h-corrected}  
& \cM^{(1)} [1_a^- 2_b^- 3_c^+ 4_h^- 5_h^+]  \supset  
 - \frac{g^3}{16 \sqrt{2} \pi^2 \mpl^2} { \langle  4  5\rangle \over [45]}
 { \langle 12\rangle^2    [35]^4 [5| p_1 p_2 |5] \over [34]^2 } 
 \nnl \times & 
\bigg\{ 
i n_+^{\cal R} C_{\cal R}f^{abc} \bigg [ G_{3gh}(s_{12},s_{13})   +   {2 \over s_{12}^3 }  \bigg ] 
+ n_-^{\cal R} d_{\cal R}^{abc} \bigg [ F_{3\gamma h}(s_{12},s_{12})   + {s_{13}- s_{23} \over s_{12}^2 s_{13} s_{23}}  \bigg ] 
\bigg \}  
.   \end{align} 
This has a smooth $1\|2$ collinear limit. 
However, the part proportional to  
$n_-^{\cal R}$ in \cref{eq:BREAK_M3g2h-corrected} develops new singularities in the $1\|3$ and $2\|3$ channels, which again can be shown to be at odds with factorization.   
The only way to save collinear factorization is to  impose the non-abelian triangle anomaly cancellation condition: 
\begin{align}  
 \sum_{\cal R} n_-^{\cal R} d_{\cal R}^{abc}  = 0 
.   \end{align}  

\vspace{1cm}

Finally, an analogous discussion allows one to also derive the mixed $U(1)$-gravitational anomaly cancellation condition.
Starting with the 3-graviton-1-photon amplitude in \cref{eq:FACT_Mhhha-mmpp}, we construct the part of  
the 5-point amplitude $\cM[1_h^- 2_h^+ 3_\gamma^+ 4_{\psi_k}^- 5_{\bar \psi_k}^+]$ due to the graviton exchange between the fermion pair and the rest of the particles. 
As before, one finds that, for $R_{3h\gamma} \to 0$, the 5-point amplitude does not have universal collinear behavior in the $2\|3$ channel.  
Adjusting the rational term as 
 \begin{align} 
   \label{eq:ACFB_Rh3gamma}
 R_{3h\gamma}[1_h^- 2_h^- 3_h^+ 4_\gamma^+]  = & 
  \bigg [ \sum_{n=1}^N  s_n  Q_n  \bigg ] 
   { \sqrt{2}  e   \over 32 \pi^2 \mpl^3 } \langle 12 \rangle^4   [34]^2  [3| p_1 p_2|3]   
  { t-u \over  s^4   } 
  \bigg [ 1    + {s^2  \over 12  t u   }  \bigg ]   
,   \end{align} 
the 5-point amplitude acquires the contribution 
      \begin{align}  
 R'[1_h^- 2_h^+ 3_\gamma^+ 4_{\psi_k}^- 5_{\bar \psi_k}^+]     = &  
\bigg [ \sum_{n=1}^N  s_n  Q_n  \bigg ] 
   { \sqrt 2 e \over   32 \pi^2 \mpl^4}  
     {  [45] \over \langle 4 5 \rangle }  
 \langle 14 \rangle^3  \langle 15 \rangle  [23]^2  [2| p_1 p_3  |2] 
 { s_{12} - s_{13}  \over s_{23}^4 } 
   \bigg [ 1    + {s_{23}^2  \over 12  s_{12} s_{13}   } 
  \bigg ]   
   ,  \end{align} 
   which makes it regular in the $2\|3$ collinear limit.  
   However, the $1\|2$ and $1\|3$ collinear limits are then irreconcilable
 with unitarity. 
 The theory violates collinear factorization unless the mixed anomaly cancellation condition 
\begin{align}
    \sum_{n=1}^N  s_n  Q_n  = 0  
\end{align}
is satisfied.  

We conclude this section by remarking that a similar analysis of amplitudes that are {\em not} proportional to an anomaly cancellation condition of course leads to no pathology in the collinear limits; an example is provided in \cref{app:INT_B}.

\section{Conclusions and Outlook} 
\label{sec:CONC} 

In this work we investigated how gauge 
anomalies are manifested in the on-shell formalism, where gauge symmetry is bypassed from the get-go. 
In this setting, gravity turns out to be a useful probe of anomalies, in the sense that certain one-loop $\mpl$-suppressed 4-point amplitudes involving photons, gluons, gravitons, and chiral fermions are proportional to the anomaly cancellation conditions. 
We construct these amplitudes with the help of unitarity methods, and use them to bootstrap higher-point amplitudes that display distinct pathologies in the presence of anomalies. 

Our main result is that in the on-shell language gauge anomalies are directly reflected in the breakdown of collinear factorization. While one-loop four-point amplitudes may exhibit non-trivial structures proportional to anomaly coefficients, they remain compatible with unitarity and collinear factorization on their own. The situation changes when we consider certain five-point amplitudes, where the singular behaviors in collinear limits cannot be reconciled with the factorization theorem in \cref{eq:FACT_Factorization-loop}. 
Another reason to consider five-point amplitudes is that collinear factorization is rigorously defined and proved for $n\geq5$. 
We showed that, in anomalous theories, the cut-constructible part of selected five-point amplitudes is at odds with collinear factorization, and attempts to restore it by adjusting rational terms inevitably introduce pathologies in other collinear limits. 
In order to avoid factorization breakdown, one has to impose the anomaly cancellation conditions
\begin{align}
\sum_n s_n Q_n^3=0, \qquad   \sum_n s_n Q_n=0, \qquad 
 \sum_{\cal R} n_-^{\cal R} d_{\cal R}^{abc}  = 0 ,
\end{align}
which correspond respectively to the vanishing of the $U(1)^3$ anomaly, the mixed $U(1)$-gravitational anomaly, and  the non-abelian triangle anomaly.

While in this work we attacked the simplest anomalous theories, it would be interesting to see how the breakdown of collinear factorization would interplay with the Green-Schwarz mechanism, namely how the addition of a new dynamical degree of freedom to the spectrum would restore broken collinear factorization of some pathological amplitude. Another interesting direction would be the study of 
anomalies of global symmetries from the on-shell perspective.  
In this case we do not expect breakdown of any fundamental principles,  
as anomalous global symmetries is perfectly allowed. 
However, the use of gravity as a probe 
might expose the physical consequences of anomalous global symmetries.
Finally, it would be interesting to rigorously prove an on-shell version of the Adler-Bardeen theorem~\cite{Adler:1969er} that anomalies are one-loop exact. 
Indeed, the collinear splitting amplitudes for gravity are also not renormalized beyond one loop~\cite{Bern:1998sv}, providing a tantalizing connection.

\section*{Acknowledgements}

We would like to thank Quentin Bonnefoy and Stefano De Angelis for illuminating discussions and for their comments on the manuscript. 

\appendix

\newpage

\section{Phase space integrals} 
\label{app:INT}

This appendix reviews the techniques for dealing with 2-body phase space integrals appearing in the calculation of one-loop massless 4-point amplitudes. 

For $s$-channel cuts, the massive internal momenta $p_X$ and $p_Y$ are decomposed as
\begin{align}
\label{eq:INT_pXYof12} 
p_X= & 
\alpha p_1+(1-\alpha) p_2-\sqrt{\alpha(1-\alpha)} \big [z w +z^{-1} \bar w \big ]
,  \notag\\
p_Y= & 
(1-\alpha) p_1+\alpha p_2+\sqrt{\alpha(1-\alpha)} \big  [z w +z^{-1} \bar w \big ]
, \end{align}
where $p_1$, $p_2$ are the incoming massless momenta, $p_1^2=p_2^2=0$.
The other two basis momenta are defined as 
$w \sigma = |2\rangle [1|$, 
$\bar w \sigma = |1\rangle [2|$, 
such that 
$p_1 w = p_2 w = p_1 \bar w = p_2 \bar w =0$,  
$w^2 = \bar w^2 = 0$, 
and 
$2 w \bar w = - s$. 
The parameter $\alpha$ is in the range $\alpha\in[0,1]$, 
and the parameter $z$ is constrained to the unit circle, $|z|=1$. 
The parametrization in \cref{eq:INT_pXYof12} is equivalent to the one in Refs.~\cite{Zwiebel:2011bx,Caron-Huot:2016cwu} with $\alpha =\cos^2 \theta$, $z=e^{i\phi}$. 
For cuts in other channels we used the analogous parametrization appropriately crossed. 
The phase space element in these variables reads 
\begin{align}
\label{eq:INT_phasespace}
\text{d}\Pi_{XY}=
\frac{\text{d}\alpha}{8\pi}
\frac{\text{d}z}{2\pi i z}
.\end{align}

Massless one-loop 4-point amplitudes can be expressed in the basis of scalar integrals defined as 
\begin{align}
\label{eq:INT_scalar-integrals}
I_\circ^s \equiv  &   
\int { \mu^{4-d} \text{d}^d k
\over i (2 \pi)^d }  
\frac{1}{k^2 (k+p_1+p_2)^2}
,\notag\\
I_\triangle^s \equiv  & 
\int { \mu^{4-d} \text{d}^d k
\over i (2 \pi)^d }
\frac{1}
{k^2(k+p_1)^2 (k+p_1+p_2)^2}
,\notag\\  
I_\Box^{st} \equiv  & 
\int { \mu^{4-d} \text{d}^d k
\over i (2 \pi)^d }
\frac{1}{k^2(k+p_1)^2(k+p_1+p_2)^2(k-p_3)^2}
, \end{align} 
where 
$d = 4 - 2 \varepsilon$. 
The remaining basis integrals $I_\circ^t$, $I_\circ^u$, 
$I_\triangle^t$, $I_\triangle^u$,
$I_\Box^{su}$, 
and $I_\Box^{tu}$ 
can be obtained from the above by crossing 
$s \leftrightarrow t$ and 
$2 \leftrightarrow 3$, 
or 
$s \leftrightarrow u$ and 
$2 \leftrightarrow 4$. 
Calculating the $s$-channel two-particle cuts of the integrals in \cref{eq:INT_scalar-integrals} using the parametrization in \cref{eq:INT_pXYof12} we find 
\begin{align} 
  \label{eq:INT_Disc2Is}
 {\rm Disc}_s   I_\circ^s = &      {i  \over 8 \pi} 
 , \nnl 
{\rm Disc}_s  I_\triangle^s = &   
{i  \over 8 \pi s }\log\epsilon 
 , \nnl  
{\rm Disc}_s I_\Box^{st}  =   & 
 {i  \over 4 \pi s t }  \bigg \{   \log\epsilon -   \log\bigg (  - {t \over s}  \bigg )     \bigg \}
      . \end{align}   
Here,  $\epsilon$ is the IR regulator defined as the cutoff of the $\alpha$ integration domain: 
$[0,1] \to [\epsilon, 1-\epsilon]$. 
It is related to the dimensional regulator  by 
 \begin{align} 
{1 \over \varepsilon} =
\log \epsilon 
+ \log \bigg ( - {s \over \mu^2 } \bigg) 
 . \end{align} 
Any $s$-channel cut of massless 4-point amplitudes can be expressed as the linear combination of the right-hand-sides in \cref{eq:INT_Disc2Is}, which allows one to easily perform  the decomposition of the result into scalar integrals. 

The analytic expressions for the scalar integrals in dimensional regularization~\cite{Ellis:2007qk} are given by
\begin{align} 
\label{eq:INT_BubbleTriangleBoxAnalytic}
I_\circ^x  = &  {c_\Gamma \over 16 \pi^2} \bigg \{  {1 \over \varepsilon} - \log \bigg ( - {x \over \mu^2 } \bigg) + 2  \bigg \} 
+ \cO(\varepsilon^0)
, \nnl 
I_\triangle^x  = &
 {c_\Gamma \over 16 \pi^2 x} \bigg \{ 
  {1 \over \varepsilon^2}  
-{1 \over  \varepsilon} 
\log \bigg ( - {x \over \mu^2 } \bigg)  
+ {1 \over 2 } \log^2\bigg ( - {x \over \mu^2 } \bigg)
  \bigg \} 
  + \cO(\varepsilon^0)
  , \nnl  
I_\Box^{xy} = &    {c_\Gamma \over 16 \pi^2 x y } \bigg \{   
{4 \over \varepsilon^2}  
- {2 \over  \varepsilon} \bigg [   \log \bigg ( - {x \over \mu^2 } \bigg)   +  \log \bigg ( - {y \over \mu^2 } \bigg)   \bigg ] 
+ 2  \log \bigg ( - {x \over \mu^2 } \bigg)   
\log \bigg ( - {y \over \mu^2 } \bigg)   
- \pi^2 
\bigg \} 
+ \cO(\varepsilon^0)
,  \end{align}  
where $c_\Gamma =  
{ (4\pi)^\varepsilon \Gamma^2(1- \varepsilon)  \Gamma(1+\varepsilon)
\over 
\Gamma(1-2\varepsilon)}$. 

\section{Factorization for non-anomalous amplitudes} 
\label{app:INT_B}

As a sanity check, in this appendix we briefly discuss an example of how collinear factorization holds for amplitudes which are not proportional to anomaly cancellation conditions.

We start with the 2-photon-2-graviton amplitude
\begin{align}
  \cM ^{(1)} \big[ 1_\gamma^- 2_\gamma^- 3_h^+ 4_h^+ \big] =& 
  \bigg[\sum_{n=1}^N Q_n^2 \bigg] \frac{ e^2 }{16 \pi^2 \mpl^2} \frac{\la12\ra^3 [34]^3}{s^3} \frac{ [13] [24]}{t} F_{2\gamma2h(s,t)}+R_{2\gamma2h}
  \notag \\ 
  F_{2\gamma2h}(s,t)=&\frac{1}{s^3}\bigg [
  \frac{2}{3} s (t-u) (t^2 - tu + u^2)  \log\bigg(\frac{t}{u}\bigg) 
  - t u ( t^2 + u^2 )\bigg( \log\bigg(\frac{t}{u}\bigg)^2 + \pi^2 \bigg)\bigg ]
  \label{eq:2 photon 2 graviton -- loop}   
, \end{align}
which is invariant under the exchange of particles 1 (3) and 2 (4) in virtue of $\frac{[13][24]}{t}=-\frac{[14][23]}{u}$.
This amplitude does not vanish in a non-chiral theory, and has nothing to do with anomalies. 
We will however utilize it for a sanity check of our method, in order to demonstrate that the factorization theorems violated by anomalous amplitudes are perfectly satisfied by non-anomalous ones. 

Consider the amplitude  $\cM[ 1_\gamma^- 2_\gamma^- 3_h^+ 4_h^+ 5_h^-]$.
For $s_{45}\to 0$, this amplitude contains a contribution from the graviton exchange  
\begin{align} 
\label{eq:BREAK_M2ph3h} 
\cM^{(1)} [1_\gamma^- 2_\gamma^- 3_h^+ 4_h^+ 5_h^-]  \supset  & 
- { \cM ^{(1)} \big[ 1_\gamma^- 2_\gamma^- 3_h^+ K_h^+ \big ] 
\cM^{(0)} [(-K)_h^-  4_h^+ 5_h^- \big]  \over s_{45} } 
=
M_{2\gamma3h} + R'_{3g2h} 
,\nnl   
 M_{2\gamma3h} = &
\bigg[\sum_n Q_n^2\bigg] \frac{e^2}{16 \pi^2 \mpl^3} { \langle  4  5\rangle \over [45]}
 {   [34]^5 \over [12]^3[35]^2 } {  [13] [24] \over s_{13} } F_{2\gamma2h}(s_{12},s_{13})
 , 
\nnl 
R'_{2\gamma3h} =  &
  { R_{2\gamma 2 h}  \over s_{45} } 
  {  \langle K 4 \rangle^6 \over  \mpl \langle K 5 \rangle^2 \langle 4 5 \rangle^2  } 
,    \end{align}
where $K = p_4 + p_5$. In the limit where $R_{2\gamma2h}$ vanishes, the 5-point amplitude in \cref{eq:BREAK_M2ph3h} displays a $1/\sqrt{s_{12}}$ singularity that cannot be fit into the factorization theorem of  \cref{eq:FACT_Factorization-loop}. Again, this pathological behavior can be eliminated by a judicious choice of rational term $R_{2\gamma2h}$, and this comes with a new term modifying the $1||3$ collinear limit. However, in this case the modification of the $1||3$ collinear limit is legit. Namely
\begin{equation}
R'_{2\gamma3h} 
\xrightarrow{1||3}\text{Split}^{(1)}_{\gamma^-}\big[ 1_\gamma^- 3_h^+ \big]
 \cM^{(0)} [P_\gamma^+ 2_\gamma^- 4_h^+ 5_h^-]+\text{Split}^{(0)}_{\gamma^+}\big[ 1_\gamma^- 3_h^+ \big]
 \cM^{(1)} [P_\gamma^- 2_\gamma^- 4_h^+ 5_h^-],
\end{equation}
and this equation can always be satisfied given a freedom in the rational term of the loop-amplitude $ \cM^{(1)} [1_\gamma^- 2_\gamma^- 3_h^+ 4_h^-] $ (which is actually purely rational).

\bibliographystyle{JHEP}
\bibliography{anomalies}

\end{document}